# Mobile phone data and COVID-19: Missing an opportunity?

Nuria Oliver[1,2], Emmanuel Letouzé[2,3], Harald Sterly[4], Sébastien Delataille[5], Marco De Nadai[6], Bruno Lepri[2,6], Renaud Lambiotte[7], Richard Benjamins[8,9], Ciro Cattuto[10,11], Vittoria Colizza[12], Nicolas de Cordes[13], Samuel P. Fraiberger[14], Till Koebe[2,15], Sune Lehmann[16], Juan Murillo[17], Alex Pentland[18], Phuong N Pham[2,19], Frédéric Pivetta[20], Albert Ali Salah[2,21], Jari Saramäki[22], Samuel V. Scarpino[23], Michele Tizzoni[11], Stefaan Verhulst[24], Patrick Vinck[2,19, 25]

March 26, 2020

[1] ELLIS, the European Laboratory for Learning and Intelligent Systems
[2] Data-Pop Alliance (USA, Colombia, Mexico, Senegal and Spain)
[3] OPAL Project
[4] University of Vienna, Austria
[5] Rosa, Belgium
[6] Fondazione Bruno Kessler, Italy
[7] University of Oxford, UK
[8] Telefonica, Spain
[9] OdiseIA, Spain
[10] University of Turin, Turin, Italy
[11] ISI Foundation, Italy
[12] INSERM, Sorbonne Université, Pierre Louis Institute of Epidemiology and Public Health, Paris, France
[13] Orange Group, France
[14] World Bank, Washington DC, USA
[15] Freie University, Berlin, Germany
[16] Technical University of Denmark, Copenhagen, Denmark
[17] Banco Bilbao Vizcaya Argentaria, Madrid, Spain
[18] Massachusetts Institute of Technology, Cambridge, USA
[19] Harvard University, Cambridge, USA
[20] Dalberg Data Insights, Belgium
[21] Utrecht University, The Netherlands
[22] Aalto University, Finland
[23] Northeastern University, Boston, MA, USA
[24] The GovLab, New York University, USA
[25] Corresponding author: pvinck@hsph.harvard.edu

## Abstract

This paper describes how mobile phone data can guide government and public health authorities in determining the best course of action to control the COVID-19 pandemic and in assessing the effectiveness of control measures such as physical distancing. It identifies key gaps and reasons why this kind of data is only scarcely used, although their value in similar epidemics has proven in a number of use cases. It presents ways to overcome these gaps and key recommendations for urgent action, most notably the establishment of mixed expert groups on national and regional level, and the inclusion and support of governments and public authorities early on.

It is authored by a group of experienced data scientists, epidemiologists, demographers and representatives of mobile network operators who jointly put their work at the service of the global effort to combat the COVID-19 pandemic.

# 1. Introduction

The Coronavirus disease 2019-2020 pandemic (COVID-19) poses unprecedented challenges for governments and societies around the world [Anderson et al., 2020]. In addition to medical measures, non-pharmaceutical measures have proven to be critical for delaying and containing the spread of the virus [Chinazzi et al., 2020; Ferguson et al., 2020; Tian et al., 2020, Zhang et al., 2020; Di Domenico et al., 2020]. This includes (aggressive) testing and tracing, bans on large gatherings, school and university closures, international and domestic mobility restrictions and physical isolation, up to total lockdowns of regions and countries. However, effective and rapid decision-making during all stages of the pandemic requires reliable and timely data not only about infections, but also about human behavior, especially on mobility and physical co-presence of people.

Seminal work on human mobility has shown that mobile phone data can assist the modeling of the geographical spread of epidemics [Bengtsson et al., 2015; Finger et al., 2016; Tizzoni et al., 2014; Wesolowski et al., 2012; Wesolowski et al., 2015]. Thus, researchers and governments have started to collaborate with private companies, most notably mobile network operators, to estimate and visualize the effectiveness of control measures. In China, Baidu data has been used to evaluate how the lockdown of Wuhan affected the spread of the virus [Kraemer et al., 2020; Lai et al., 2020]. In Italy, researchers and local governments are collaborating to estimate the effectiveness of travel restrictions [Pepe et al., 2020; adnkronos, 2020]. European authorities (including Austria, Belgium, Germany, Italy, France, and Spain) are working with researchers and mobile network operators to understand the compliance and impact of the social distancing measures put in effect to combat the COVID-19 pandemic and to identify and predict potential hotspots of the disease [El País 2020; The Verge 2020]. Using the city of Boston, United States, as a test case, researchers aggregated location data from over 180 apps to enable precise measurements of social distancing on a day-by-day basis, and project in great detail the effects of different policies on the spread of COVID-19 [Martin-Calvò et al., 2020].

However, there is currently hardly any coordination or information exchange between these national or even regional initiatives, as also highlighted by Buckee *et al.* in a letter to Science [Buckee et al., 2020]. Although ad-hoc mechanisms can be effectively (but not easily) developed at the local or national level, regional or even global collaborations seem all but impossible given the number of actors, the range of interests and priorities, the variety of legislations concerned, and the need to protect civil liberties. The global scale and spread of the COVID-19 pandemic highlight the need for a more harmonized or coordinated approach.

In the following sections, we outline in what ways and how mobile phone data can help to better target and design measures to contain and slow the spread of the COVID-19 pandemic. We identify the key reasons why this is not happening on a much broader scale, and we give recommendations on how to make mobile phone data work against the virus.

## 2. How can mobile phone data help to tackle the COVID-19 pandemic?

The use of mobile phone data into analytical efforts to control the COVID-19 pandemic can offer a critical contribution to four broad areas of investigations:

- **Situational awareness** would benefit from increased access to previously unavailable population estimates and mobility information to enable stakeholders across sectors better understand COVID-19 trends and geographic distribution.
- **Cause-and-effect** use cases can help stakeholders identify the key drivers and consequences of implementing different measures to contain the spread of COVID-19. They aim to establish which variables make a difference for a problem and whether further issues might be caused.
- **Prediction** tasks would leverage real-time population counts and mobility data to enable new predictive capabilities and allow stakeholders to assess future risks, needs, and opportunities.
- **Impact assessment** aims to determine which, whether, and how various interventions affect the spread of COVID-19 and requires data to identify the obstacles hampering the achievement of certain objectives or the success of particular interventions.

| Situational awareness | Cause-and-effect |
|---|---|
| • What are the most common mobility flows within and between COVID-affected cities and regions?<br>• Which areas are spreading the epidemics acting as origin nodes in a mobility network, and thus could be placed under mobility restrictions?<br>• Are people continuing to travel or congregate after social distancing and travel restrictions were put into place?<br>• Are there hotspots at higher risk of contamination (due to a higher level of mobility, higher concentration of population)?<br>• What are the key entry points, locations and movements of roamers/tourists? | • What are variables that determine the success of social distancing approaches?<br>• How do local mobility patterns impact the burden on the medical system?<br>• Are business' social distancing recommendations resulting in more workers working from home?<br>• In what sectors are people working most from home?<br>• What are the social and economic consequences of movement restriction measures? |
| **Prediction** | **Impact** |
| • How are certain human mobility patterns likely to affect the spread of coronavirus? And what is the likely spread of COVID-19, based on existing disease models and up-to-date mobility data?<br>• What are the likely effects of mobility restrictions on children's education outcomes?<br>• What are likely to be the economic consequences of restricted mobility for businesses? | • How have travel restrictions impacted human mobility behavior and likely disease transmission?<br>• What is the potential of various restriction measures to avert infection cases and save lives?<br>• What is the effect of mandatory social distancing measures, including closure of schools?<br>• How has the dissemination of public safety information and voluntary guidance impacted human mobility behavior and disease spread? |

Table 1: Examples of questions by areas of investigation

**Mobile phone data in different epidemiological phases**

The importance and relevance of these areas of inquiry differ at various stages of the outbreak, but mobile phone data provide value throughout the epidemiological cycle, as shown in Fig.1.

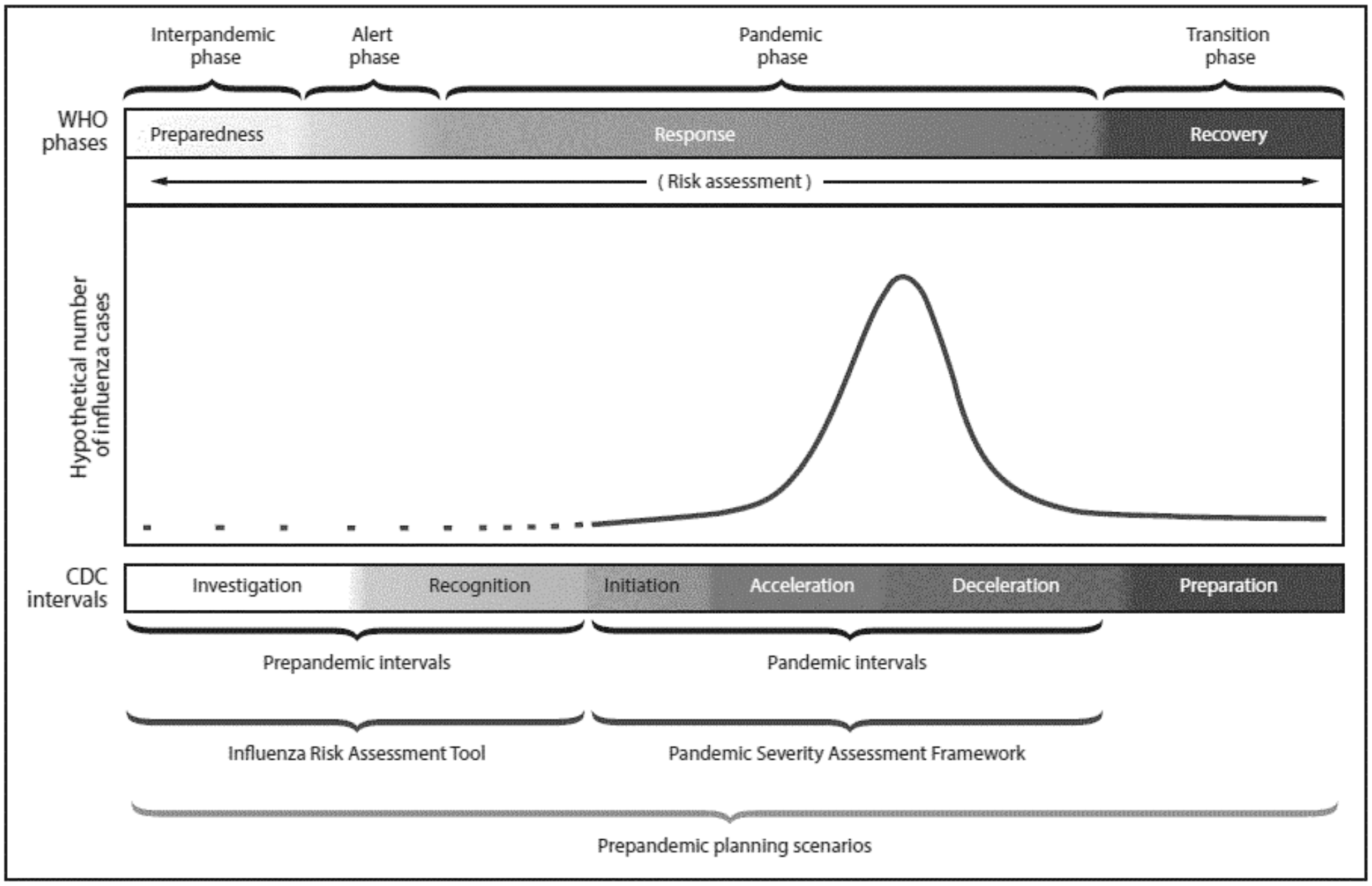

Figure 1: Pandemic intervals as defined by the US Center for Disease Control [CDC 2014]

- In the early **recognition and initiation phase** of the pandemic, the focus is on situational analysis and the fast detection of infected cases and their contacts. Research has shown that quarantine measures of infected individuals and their family members, combined with surveillance and standard testing procedures, are effective as control measures in the early stages of the pandemic [Koo et al, 2020]. Individual mobility and contact data offer information about infected individuals, their locations and social network. Contact data can be collected through mobile apps [Dong et al., 2019; Ferretti et al., 2020], interviews, surveys, voluntary sharing of data, or by accessing individual mobile phone data, which presents serious ethical and legal concerns.
- During the **acceleration phase**, when community transmission reaches exponential levels, the focus is on interventions for containment, which typically involve social contact and mobility restrictions. Aggregated mobile phone data is here crucial to assess the efficacy of implemented policies through the monitoring of mobility between and within affected municipalities. Mobility information also contributes to the building of more accurate epidemiological models that can explain and anticipate the spread of the disease, as shown for Influenza A (H1N1) outbreak [Balcan et al., 2009]. These models, in turn, can inform the mobilization of resources (e.g. respirators, intensive care units).

- Finally, during the **deceleration and preparation phases**, as the peak of infections is reached, restrictions will likely be lifted. Continued situational monitoring will be important as the COVID-19 pandemic is expected to come in waves [Ferguson et al. 2020, see Fig. 2]. Near real-time data on mobility and hotspots will be important to understand how lifting and re-establishing various measures translate into behavior, especially to find the optimal combination of measures at the right time (e.g. general mobility restrictions, school closures, banning of large gatherings), and to balance these restrictions with aspects of economic vitality. After the pandemic has subsided, mobile data will be helpful for post-hoc analysis of the impact of different interventions on the progression of the disease, and cost-benefit analysis of mobility restrictions. The experience may also help design technology such as the Korean app (*Corona 100m*) to help people re-starting a (careful) social life without too much stress and further minimizing the spread of a disease. Along this line, researchers at the Massachusetts Institute of Technology and other collaborators are working on *Private Kit: Safe Paths* [Barbar et al., 2020], a free open-source and privacy-first contact-tracing technology that provides individuals with information on their interaction with diagnosed COVID-19 carriers.

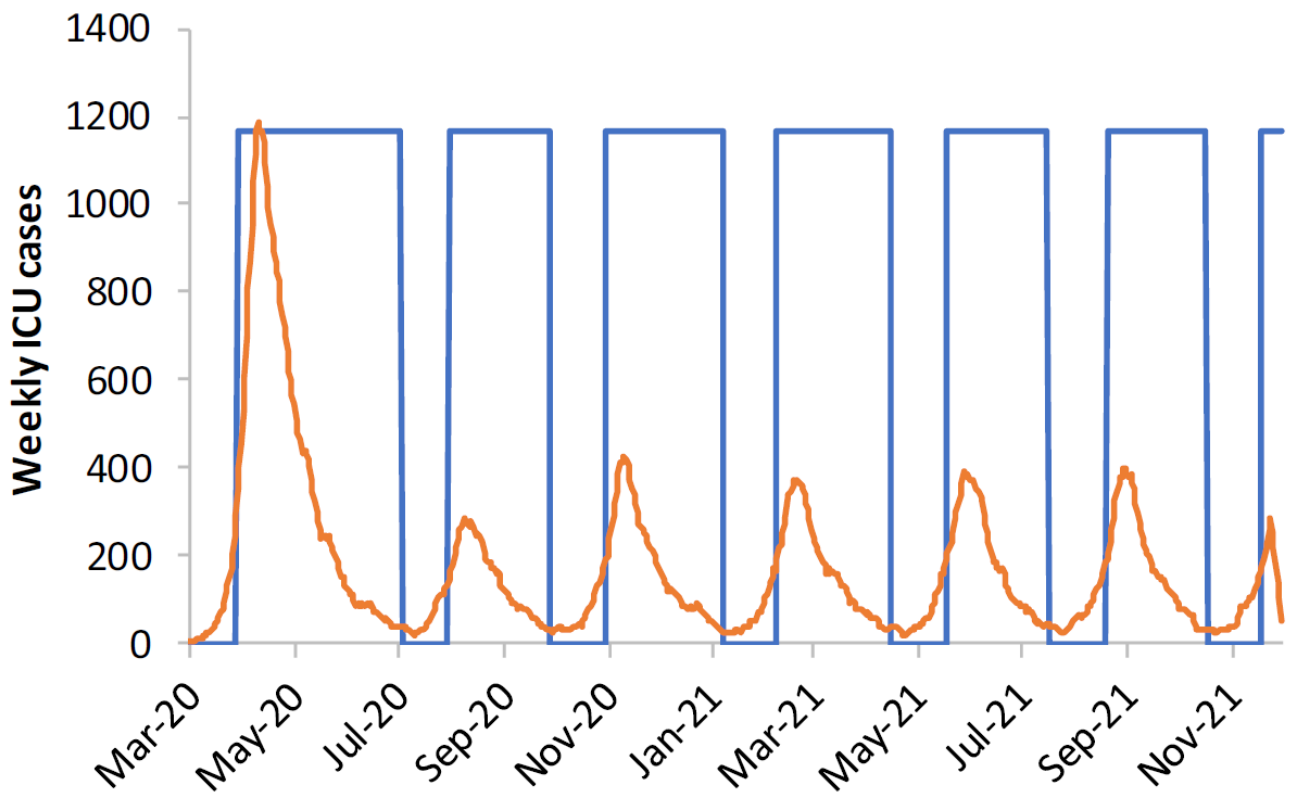

Fig 2: Illustration of outbreaks (orange curve) and adaptive triggering (blue curve) of suppression strategies. [Ferguson et al. 2020: 12]

## 3. Specific metrics for data-supported decisions

Mobile phone traces can be exploited to infer human mobility and social interactions. Researchers have used various types of mobile network data (e.g. Call Details Records, x Data Records and passive records), GPS and mobile application data. Some of this data, such as CDRs, are collected by mobile network operators for billing reasons, and each record contains information about the time and (approximate) geographical location of customers' interactions with the phone. Other data consists of very accurate geographic information that can be collected at a large scale by smartphone applications (Walle 2020), however often with limited coverage in developing countries and raising complex questions around privacy and civil liberties.

We here identify some examples of aggregated metrics that can be computed from such data sources:

- **Origin-destination (OD) matrices** are especially useful in the first epidemiological phases, where the focus is to assess the mobility of people. Researchers can compute the number of people that move from two different areas daily, which is assumed to be a proxy of human mobility. The areas might be identified as municipalities, provinces or even regions. They could be compared to a reference period in order to assess the reduction in mobility and could inform spatially explicit disease transmission models to evaluate the potential benefit of such reductions. They are also useful to monitor the impact of different social and mobility contention measures and to identify regions where the measures might not be effective or followed by the population.
- **Dwell estimations and hotspots** are estimates of particularly high concentration of people in an area, which can be favorable to the transmission of the virus. These metrics are typically constructed within a municipality by dividing the city into grids or neighborhoods [Louail et al., 2014]. The estimated number of people in each geographical unit can be computed with different time granularities (e.g. 15', 60', 24h).
- **Amount of time spent at home, at work, other** are estimates of the individual percentage of time spent at home/work/other location, which can be useful to assess the local aversion to countermeasures adopted by governments. The home and work locations need to be computed in a period of time prior to the deployment of mobility restrictions measures. The percentage of time spent in each location needs to be computed from stationary points, thus excluding people that move. Variations of the time spent on different locations is generally computed on an individual basis, and then spatially aggregated.
- **Contact matrices** estimate the number and intensity of the face-to-face interactions people have in a day. They are typically computed by age-groups, and they showed to be extremely useful to assess and determine the decrease of the reproduction number of the virus [Zhang et al., 2020]. However, it is still challenging to estimate face-to-face interactions from co-location data [Génois and Barrat 2018].

Although there is still little information about the age-specific susceptibility to COVID-19 infection, it is clear that age is an important risk factor for COVID-19 severity. We highlight, therefore, the importance of estimating the metrics mentioned above by age groups [Zhang et al., 2020].

# 4. Why is the use of mobile phone data not widespread, or a standard, in tackling epidemics?

At present, there is still a scarce strategic use of mobile phone data for tackling the COVID-19 pandemic. Although local alliances are starting to be formed, an internationally concerted action is still missing.

In the past, notably in the Ebola outbreak of 2014-2016, several pilot or one-off activities have been initiated. However, there was no transition to "business as usual" in terms of standardized procedures to leverage mobile phone data or establish mechanisms for 'data readiness' in the country contexts [The Economist 2014, McDonald, 2016]. Technology has evolved (e.g. OPAL, Flowkit, also commercial products), high-level meetings have been held (e.g. the European Commission's B2G Data Sharing high-level expert group), data analysis and sharing showed promising results, but the use of metrics derived from mobile phone data by governments and local authorities is almost *non-existent* [Maxmen, 2019]. Why does this gap persist despite technical progress?

We identify the following five reasons:

- **Capacity, awareness and digital mindset of governments and public authorities:** Public authorities frequently lack the capacity, both for processing information that often is complex and requires multidisciplinary expertise (e.g. mixing location and health data, specialized modeling), as well as for establishing the necessary interdisciplinary teams and collaborations. Also, many government units are understaffed and sometimes also lack technological equipment. During the COVID-19 pandemic, most authorities are overwhelmed by the multiplicity and simultaneity of requests; as they have never been confronted with such a crisis, there are few predefined procedures and guides, so targeted and preventive action is quickly abandoned for mass actions. Finally, many public authorities and decision-makers are not aware of the value that mobile phone data would provide for decision making and are often used to make decisions without knowing the full facts and under conditions of uncertainty.
- **Access to data:** Most companies, including mobile network operators, tend to be very reluctant to make data available --- even aggregated and anonymized --- to researchers and/or governments. Apart from data protection issues, such data are also seen and used as commercial assets, thus limiting the potential use for humanitarian goals if there are no sustainable models to support operational systems. One should also be aware that not all mobile network operators in the world are equal in terms of data maturity.

Some are actively sharing data as a business, while others have hardly started to collect and use data.

- **Concerns about privacy and data protection.** Governments in China, South Korea, Israel and elsewhere have openly accessed and used personal mobile phone data for tracking individual movements and for notifying individuals. However, in other regions, such as in Europe, both national and regional legal regulations limit such use (especially the European Union law on data protection and privacy known as the General Data Protection Regulation - GDPR). Furthermore, around the world, public opinion surveys, social media and a broad range of civil society actors including consumer groups and human rights organizations have raised legitimate concerns around the ethics, potential loss of privacy and long-term impact on civil liberties resulting from the use of individual mobile data to monitor COVID-19 and send personalized notifications to citizens.
- **Researchers and domain experts** tend to define the scope and direction of analytical problems from their perspective and not necessarily from the perspective of governments' needs. Critical decisions have to be taken, while key results are often published in scientific journals and in jargon that are not easily accessible to 'outsiders', including government workers and policy makers.
- There is little to **no preparedness for immediate** and rapid action. On country levels, there are too few latent/'standing' mixed teams, composed of a) representatives of governments and public authorities, b) mobile network operators, c) different topic experts (virologists, epidemiologists, data analysts); and there are no procedures and protocols predefined.

## 5. From words to deeds: A call to action to fight COVID-19

To effectively build the best, most up to date, most relevant, and actionable knowledge, we call on governments, mobile network operators and researchers to form mixed teams:

- Governments should be aware of the value of information and knowledge that can be derived from mobile phone data analysis, especially for sensible targeting and monitoring the necessary measures to contain the pandemic. They should enable and leverage the fair and responsible provision/use of aggregated and anonymized data for this purpose.
- Mobile network operators should take their social responsibility and the vital role that they can play in tackling the pandemic. They should reach out to governments and the research community.
- Researchers and domain experts (e.g. virologists, epidemiologists, demographers, data scientists, computer scientists and computational social scientists) should acknowledge interdisciplinarity and context specificities and sensitivities. They need to include governments and public authorities early on and throughout their efforts to identify the most relevant questions and knowledge needs. Creating multi-disciplinary inter-institutional teams is of paramount importance.

The following principles would significantly improve the effectiveness of these mixed teams:

- **Include governments.** Relevant government and public authorities should be involved early, and researchers need to build upon their knowledge systems and need for information. One key challenge is to make insights actionable–how can findings such as propagation maps finally be utilized (e.g. for setting quarantine zones, informing local governments, targeting communication). At the same time, expectations must be realistic: decisions on measures should be based on facts, but are in the end, always political decisions.
- **Liaise with data protection authorities and civil liberties advocates early on, transparently**, and have quick iteration cycles with them. Consider creating an ethics and privacy advisory committees to oversee and provide feedback on the projects. This ensures that privacy is maintained and raises potential user acceptance. It is possible to make use of aggregated mobile phone data in line with even the relatively strict European regulations (GDPR). Earlier initiatives have established principles and methods for sharing data or indicators without endangering any personal information and build privacy-preserving solutions that use only incentives to manage behavior [Buckee et al., 2018; de Montjoye et al., 2018; Oliver et al., 2019]. The early inclusion of the data protection authority in Belgium (see box) has led to the publishing of a statement by the European Data Protection Board on how to process mobile phone data in the fight against COVID-19 [EDPB 2020]. Even while acknowledging the value of mobile phone data, the urgency of the situation should not lead to losses of data privacy and other civil liberties that might become permanent after the pandemic. In this regard, the donation of data for good (e.g. the MIT app) and the direct and limited (in time and scope) sharing of aggregated data by mobile network operators with (democratic) governments and researchers seems to be less problematic than the use of individual location data commercially acquired, brought together and analyzed by commercial enterprises. More generally, any emergency data system set to monitor COVID-19 and beyond must follow a balanced and well-articulated set of data policies and guidelines and be subjected to risk assessments.
- **Exchange internationally**, with other domain experts, but also with other initiatives and groups; share findings quickly – there will be time for peer-reviewed publications later. Especially in countries with weaker health systems, the targeting and effectiveness of non-pharmaceutical interventions might make a big difference. This might also imply that relevant findings are translated from English to other relevant languages.
- **Prepare for all stages of the pandemic.** For later stages of the pandemic, and for the future, a minimum level of "preparedness" for immediate and rapid action is needed. On country levels, there will be a need of 'standing' mixed teams; basic agreements and legal prescriptions should be in place, procedures and protocols predefined (also for "appropriate anonymization and aggregation protocols" [Buckee et al., 2018]).

Finally, in addition to (horizontal) international exchange, we also need international approaches that are coordinated by supranational bodies. National initiatives might help to a certain extent but will not be sufficient in the long run. A global pandemic necessitates globally or at least regionally coordinated work. Here, promising approaches are emerging: the EU Commission has on 23.03.2020 called upon European mobile network operators to hand over anonymized and aggregated data to the Commission to track virus spread and determine priority areas for medical supplies [Politico 2020], while other coordination initiatives are emerging in Africa, Latin America and the Mena-Region. It will be important for such initiatives to link up, share knowledge and collaborate.

**Belgium: “Data Against Corona” taskforce.**

On March 13th 2020, the Minister of Health and the Minister of Telecom and Privacy kicked-off a “Data Against Corona” taskforce that gathered four types of actors: i) representatives from the ministries, ii) data providers such as Mobile Network Operators and Sciensano (i.e. the national health data provider), iii) an operational team of experts (e.g. data science, spatial epidemiology, digital entrepreneurs), and iv) privacy and ethics committees.

The goal of the taskforce was to flatten the curve of the epidemic, by combining location and epidemiological data, for targeted use cases (e.g. the Minister of Health needs to monitor the reduction in mobility following national confinement measures, needs to predict where the virus is likely to propagate to allocate resources). The Minister of Health’s teams prioritized the issues that would be most helpful, and the task force then collected and analyzed the relevant data and formulated the results as actionable information.

To effectively support public authorities, the taskforce was built on these success factors: 1) A clear mandate from the government, with active participation from the cabinets; 2) a centralized, multi-disciplinary team that had proven experience from other epidemics (e.g. Ebola, Zika) and could leverage their models in an agile manner; 3) a legal and technical framework that guarantees privacy rights are preserved (the Belgian Data Protection Authority pre-authorized the analyses to be done before data processing activities even have begun); and 4) feedback loops where public authorities continuously ranked the use cases by importance, confirmed which information needed to be provided daily, and linked the results with other task forces supporting the national crisis cell (e.g. resource allocation taskforce).

In the Belgian example, public authorities took a “privacy-first” approach: use cases were ranked on effectiveness and privacy impact. From the resulting matrix, the taskforce prioritized a first wave of use cases based on anonymized and aggregated output. The taskforce monitored the evolution of the disease closely and prepared the grounds for use cases that require a stronger trade-off of effectiveness over privacy (i.e. data processing of higher granularity data).

A similar task force was created in Spain on March 23rd, starting with a pilot in the Valencian region, to be further extended to the rest of the territory.